\documentclass[fleqn,10pt]{wlscirep}

\title{Gravity measurements below $10^{-9}$ $g$  with a transportable absolute quantum gravimeter}

\author[1]{Vincent M{\'e}noret}
\author[1]{Pierre Vermeulen}
\author[2]{Nicolas Le Moigne}
\author[3]{Sylvain Bonvalot}
\author[4]{Philippe Bouyer}
\author[5]{Arnaud Landragin}
\author[1,*]{Bruno Desruelle}
\affil[1]{MUQUANS, Institut d'Optique d'Aquitaine, rue Fran{\c c}ois Mitterrand, 33400, Talence, France}
\affil[2]{G{\'e}osciences Montpellier, CNRS, Universit{\'e} Montpellier, UA, F--34095 Montpellier, France}
\affil[3]{GET, IRD, CNRS, CNES, Universit{\'e} de Toulouse, F--31400 Toulouse, France}
\affil[4]{LP2N, Laboratoire de Photonique Num{\'e}rique et Nanosciences, Institut d'Optique Graduate School, rue Fran{\c c}ois Mitterrand, 33400 Talence, France}
\affil[5]{LNE--SYRTE, Observatoire de Paris, Université PSL, CNRS, Sorbonne Universit{\'e}, 61 avenue de l'Observatoire, F--75014 Paris, France}

\affil[*]{bruno.desruelle@muquans.com}

\begin{abstract}
Gravimetry is a well-established technique for the determination of sub-surface mass distribution needed in several fields of geoscience, and various types of gravimeters have been developed over the last 50 years. Among them, quantum gravimeters based on atom interferometry have shown top-level performance in terms of sensitivity, long-term stability and accuracy. Nevertheless, they have remained confined to laboratories due to their complex operation and high sensitivity to the external environment. Here we report on a novel, transportable, quantum gravimeter that can be operated under real world conditions by non-specialists, and measure the absolute gravitational acceleration continuously with a long-term stability below 10~nm.s$^{-2}$ (1~$\mu$Gal). It features several technological innovations that allow for high-precision gravity measurements, while keeping the instrument light and small enough for field measurements. The instrument was characterized in detail and its stability was evaluated during a month-long measurement campaign.
\end{abstract}
\begin{document}

\flushbottom
\maketitle

\thispagestyle{empty}

\section*{Introduction}
Over the past few decades, gravimetry has proven a powerful tool for geoscience. Its potential in many different fields has been discussed in detail\cite{van_camp_2017}. The value of gravity at the Earth's surface is directly related to sub-surface mass distribution, and the analysis of both the temporal and spatial variations of the gravitational field has allowed for the characterization of several geophysical phenomena -- including ice mass changes\cite{makinen_2007,van_dam_2017}, the monitoring of volcanoes\cite{carbone_2017} and ground water resources \cite{kennedy_2016,fores_2017}, the study of subsidence in low-lying areas\cite{van_camp_2011}, the monitoring of geothermal reservoirs\cite{pearson_2017} and the detection of underground cavities\cite{romaides_2001}.

Earth's gravitational acceleration $g$ varies roughly between 9.78~m.s$^{-2}$ and 9.83~m.s$^{-2}$ over the whole Earth. The daily fluctuation, induced by the deformation of the planet by tides, is about $10^{-7}~g$. The variations of $g$ investigated in geoscience are usually at a smaller level and the level of relative precision required for an operational instrument is of the order of one part per billion, or 10~nm.s$^{-2}$ (1~$\mu$Gal). Several technological solutions have been developed to reach this demanding requirement, and the instruments used over the past 50 years were reviewed by Van Camp et al\cite{van_camp_2017}. Absolute gravimeters yield an accurate value of $g$ and are necessary to calibrate relative instruments and measure their drift. They are usually based on the measurement of the distance traveled by a free-falling corner-cube reflector in a vacuum chamber by laser interferometry  \cite{niebauer_1995}. While such absolute gravimeters can nowadays be operated in the field, the technology still makes it difficult to reach both the best operability and sensitivity. In particular, these instruments have moving mechanical parts that make them unsuitable for long-term continuous measurements.

Absolute measurements at the level of 10~nm.s$^{-2}$ have also been demonstrated with gravimeters based on atom interferometry\cite{kasevich_1991,peters_1999,pereira_2016}, which rely on the wave nature of matter postulated by quantum mechanics. Similar to classical absolute gravimeters, they measure the acceleration of free-fall test masses (in this case cold atoms) compared to the local ground reference frame. Atomic gravimeters have already demonstrated top-level performance in terms of sensitivity, long-term stability and accuracy in international comparisons \cite{gillot_2014,farah_2014,freier_2016,zhou_2015,jiang_2012,francis_2013,wang_2018}, and several demonstrations have shown promise for in-field and onboard applications \cite{freier_2016,bidel_2013,wu_2014,bidel_2018}. However, these instruments were not practically suited for geophysical surveys because of their limited transportability and high complexity.

Here, we report on both operability and sensitivity at the level of 10~nm.s$^{-2}$ with an atomic gravimeter. Our novel Absolute Quantum Gravimeter (AQG) includes conceptual and technological developments that made it possible to reach compactness, transportability, low maintenance and non-expert operation, both for continuous observatory measurements and gravity mapping. These developments include a hollow pyramid reflector \cite{bodart_2010,bouyer_2010} and a compact telecom based laser system. The resulting sensor measures gravity at a 2~Hz repetition rate with a sensitivity of 500~nm.s$^{-2}.\mathrm{Hz}^{-1/2}$ and a long term stability below 10~nm.s$^{-2}$ with a short installation and warm-up time. Combined, these features represent a significant technological step forward, and have enabled the first long-term measurement campaign with a quantum gravimeter in a geophysical observatory, which comprised several weeks of continuous gravity measurements on the Larzac plateau in France.

\section*{The Absolute Quantum Gravimeter}

\subsection*{Measurement principle}
The AQG measurement sequence is based on matter-wave interferometry with Rubidium atoms using two-photon stimulated Raman transitions\cite{kasevich_1991}. This type of sequence has been extensively studied and used for precision measurements\cite{peters_1999,gillot_2014,farah_2014,freier_2016}. Here, we give a brief description of the measurement principle; the step-by-step procedure can be found in the Supplementary Methods.

The sequence consists of three counterpropagating Raman pulses of duration 10, 20 and 10~$\mu$s in a $\pi/2 - \pi - \pi/2$ configuration. Between these pulses, the the atoms are in near-perfect free fall for an interrogation time of $T=60$~ms. The output ports of the interferometer are labeled by the internal states $\lvert 5^2S_{1/2}, F=1, m_\mathrm{F}=0 \rangle$ and  $\lvert 5^2S_{1/2}, F=2, m_\mathrm{F}=0 \rangle$ of the atom \cite{borde_1989}. Fluorescence detection is used to count the number of atoms in each level and measure the interferometric phase shift. For cooling, Raman and fluorescence detection, the lasers are tuned close to the D$_2$ line of $^{87}$Rb, with a wavelength $\lambda \approx 780$~nm. The proportion of atoms in the $F=2$ state is given by
\begin{equation}
P =0.5 \times \left( 1 - C \cos \Phi \right)
\end{equation}
where $C$ is the contrast of the fringes and $\Phi$ the interferometric phase shift. The contrast of the fringes is lower than 1, mainly because of velocity selection effects during the Raman pulses \cite{kasevich_1991b}. The phase shift $\Phi$ is given by
\begin{equation}
\Phi = (k_\mathrm{eff}g - 2\pi\alpha)T^2,
\end{equation}
where $k_\mathrm{eff} = 4\pi/\lambda \approx 16 \times 10^6$~m$^{-1}$ is the effective wavevector of the two-photon transition and $\alpha \approx 25$~MHz.s$^{-1}$ is a frequency chirp applied to the Raman lasers to compensate the Doppler effect. We operate the instrument around the null phase shift by servo-locking the frequency chirp $\alpha$ on the detection ratio $P$ so as to constantly maintain $k_\mathrm{eff}g - 2\pi\alpha = 0$ and stay on the central fringe of the interferometer \cite{louchet_2011}. From there we derive
\begin{equation}
g = 2\pi\frac{\alpha}{k_\mathrm{eff}}.
\label{eq:g}
\end{equation}

The sensitivity of the instrument is limited both by the signal-to-noise ratio of the detection and by the contrast. A contrast $C$ of 40\% and a detection signal-to-noise ratio of 150 correspond to an effective signal-to-noise ratio (SNR) of $0.4 \times 150 = 60$. At mid-fringe, the sensitivity is given by
\begin{equation}
\frac{\delta g}{g} = \frac{1}{k_\mathrm{eff}gT^2 \mathrm{SNR}}.
\end{equation}
With the previous parameters, $\delta g/g\approx 3 \times 10^{-8}$. This constitutes the single-shot sensitivity floor of the instrument, but this quantity can be deteriorated by other factors, such as laser phase noise or vibrations\cite{farah_2014}. However, since the instrument's repetition rate is on the order of 2~Hz, the performance is improved by averaging over time to reach a long-term relative stability close to $1 \times 10^{-9}$.

\subsection*{Instrument description}
\begin{figure}[ht]
\centering
\includegraphics[width=\linewidth]{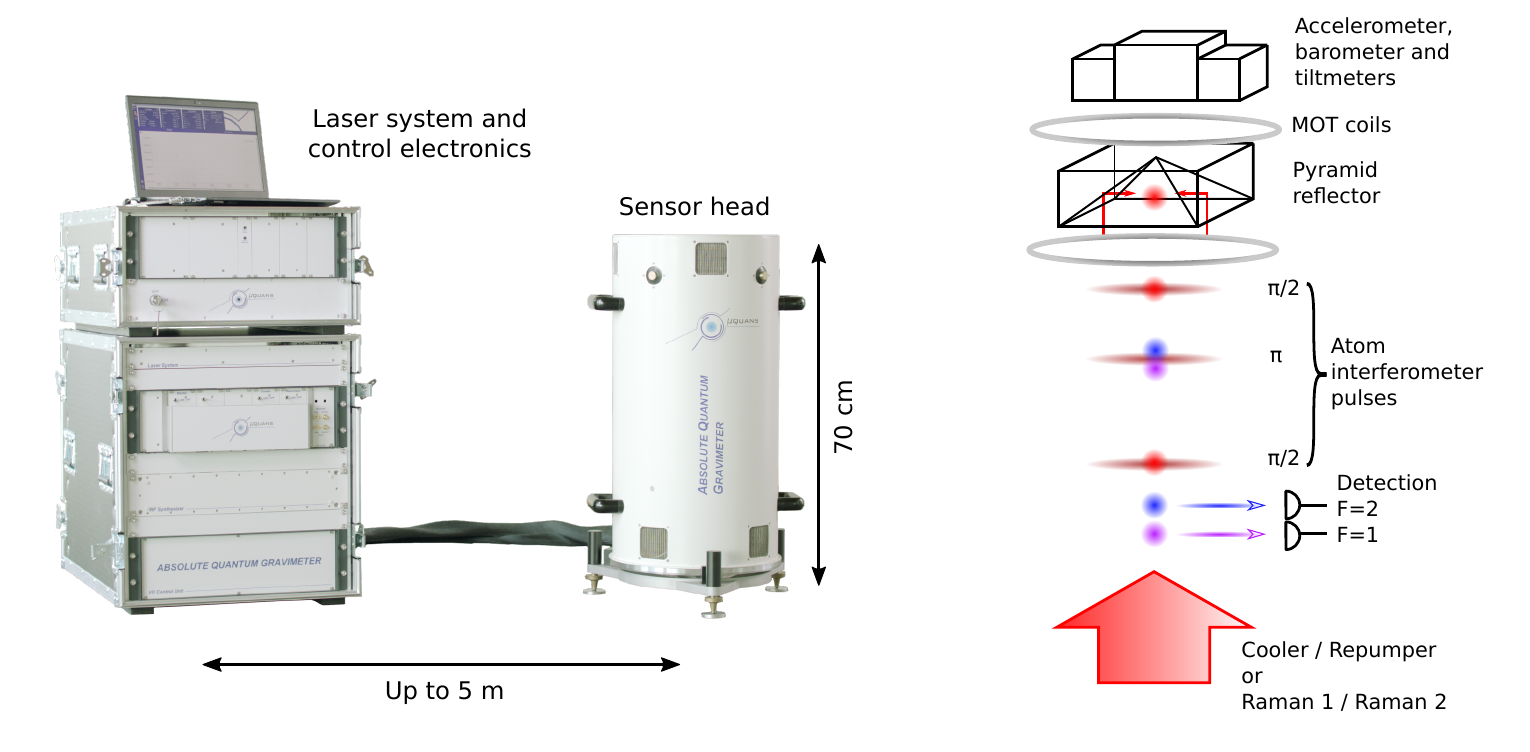}
\caption{Left: picture of the first Absolute Quantum Gravimeter (AQG-A01). The system can be setup at a measurement location in less than 20~min, warm-up time is of the order of 1~h.  The sensor head weighs approximately 30~kg.  It is mounted on an adjustable tripod with precision leveling titanium-tipped feet that can be adapted to various terrains. The measurement height is 55~cm. Right: sketch of the sensor head and measurement principle\cite{bodart_2010}. Approximately $10^7$~atoms are loaded in a magneto-optical trap (MOT) inside the pyramid and cooled down to 2~$\mu \mathrm{K}$. The $\pi/2 - \pi - \pi/2$ atom interferometry sequence is performed once the atoms are in free-fall. At the end of the sequence, we collect fluorescence on a set of photodiodes and compute the proportion of atoms in each output port of the interferometer, labeled by their internal state. A high-precision accelerometer is attached to the top of the vacuum chamber, as close as possible to the pyramidal reflector. Its signal is used to apply a real-time correction to the laser phase, in order to reject seismic noise. Two tiltmeters and a barometer are also attached to the sensor to ensure high accuracy and long-term stability of the gravity measurement.}
\label{fig:visuel_AQG}
\end{figure}

The AQG is made of two sub-units, the sensor head and the control system (fig. \ref{fig:visuel_AQG}). The sensor head houses the vacuum chamber where the measurement of gravity is performed. It can easily be set up at the measurement location, and is separated from the laser system and control electronics by a 5~m set of cables, which includes an optical fiber. The instrument can be installed in less than 20~min and is ready to measure after 1~h of warm-up time. The only adjustment to be performed by the operator during the installation process is setting the verticality of the sensor head. The angle is measured by two tiltmeters and the software indicates how the leveling tripod screws should be turned. Using this procedure, an operator can reach a verticality error of approximately 100~$\mu$rad. When not in use, the gravimeter can be completely powered off for several weeks with no significant impact on its vacuum level or warm-up time.

Using a pyramidal reflector and a single-beam geometry, we load $10^7$~$^{87}$Rb atoms in a magneto-optical trap (MOT) in 250~ms and cool them down below 2~$\mu$K in an optical molasses. We then select the atoms in the $\vert F = 1, m_F = 0 \rangle$ state by using a microwave selection. After approximately 30~ms (4.4~mm) of free-fall, the atoms exit the pyramidal reflector and we apply the $\pi/2 - \pi - \pi/2$ sequence of Raman laser pulses with the parameters described above ($\tau=10$~$\mu$s, $T=60$~ms, $C = 40$\%). The interferometer pulses are delivered by the same laser as the one used for cooling and detection, the compact configuration with the pyramidal reflector allows the use of a single beam as demonstrated by Bodart et al\cite{bodart_2010}. At the bottom of the chamber (150~mm below the position of the trap), we measure the proportion of atoms in each output port of the interferometer with a detection signal-to-noise ratio of 150. We lock the frequency chirp to the atomic signal in order to stay on the central fringe of the interferometer and derive $g$ from the values of $\alpha$ and $k_\mathrm{eff}$. To measure gravity at the $10^{-9}$ level, both of these parameters must be known with an uncertainty lower than 1 part in $10^9$. We also periodically reverse the orientation of the effective Raman wavevector to reject the systematic effects that do not depend on the sign of $k_\mathrm{eff}$ \cite{louchet_2011}.

\subsection*{Laser system}
The stability and performance of the laser system and control electronics are crucial for both the quality of the gravity measurement and the way the instrument is operated. In particular, the spectral properties of the laser will have an influence on the signal-to-noise ratio and long-term stability of the measurement. Its size, weight, robustness and ease of use determine the level of transportability and convenience for a daily operation of the gravimeter.

Our laser system complies with these constraints by using a frequency-doubled telecom solution \cite{leveque_2014,theron_2015}. Lasers operating in the telecom C-band around 1560~nm are amplified and converted to 780~nm in nonlinear crystals. The laser system is completely fibered, making it insensitive to misalignments and vibrations. Our architecture is based on a fixed-frequency reference laser, and two independent tunable slave lasers.

A master reference laser diode, emitting at 1560~nm, is frequency-doubled in a periodically-poled lithium niobate (PPLN) waveguide crystal and frequency locked to the $F=3$ to $F=4$ crossover transition of $^{85}\mathrm{Rb}$ using saturated absorption spectroscopy. Two slave lasers, used for both cooling and Raman excitation are frequency offset-locked to the master laser. Their frequency can be tuned up to 1~GHz in 200~$\mu$s. The first slave laser addresses the $\lvert 5^2S_{1/2}, F=1 \rangle$ level (repumping and Raman 1) and the second one is near-resonant with the $\lvert 5^2S_{1/2}, F=2 \rangle$ level (cooling and Raman 2). At the beginning of the Raman interferometer, the lasers are detuned by approximately 700~MHz and the frequency lock loop of the cooling and Raman 2 laser is switched to a phase lock loop on the Raman 1 laser, which remains frequency-locked to the reference laser. The phase difference between the two Raman lasers has a direct influence on the phase shift $\Phi$ of the atom interferometer \cite{kasevich_1991,cheinet_2008}. The phase lock loop is therefore necessary to ensure that that the level of residual phase noise does not deteriorate the sensitivity of the gravimeter.

\begin{figure}[ht]
\centering
\includegraphics[width=\linewidth]{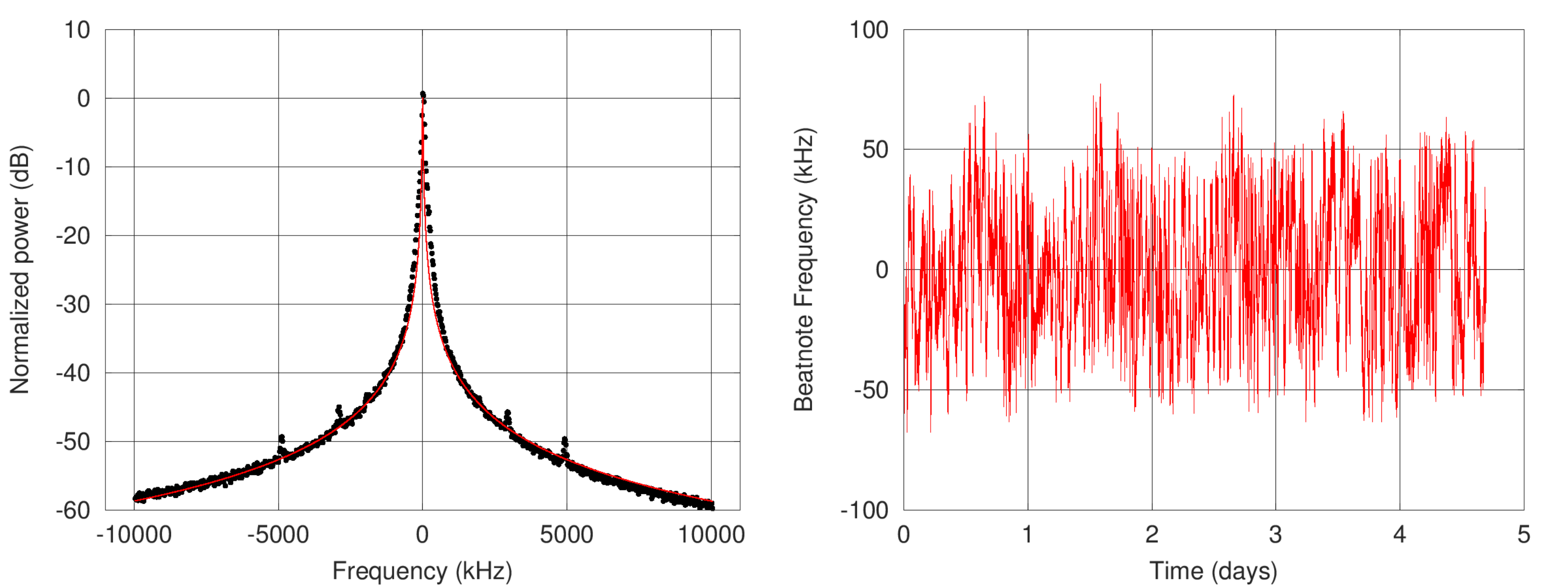}
\caption{Laser linewidth (left) and long-term frequency stability measurements (right). We make a beatnote between the laser of the gravimeter and a similar independent laser. A Lorentzian fit of the tails of the beatnote gives a linewidth of less than 12~kHz for each laser. By recording the frequency of the beatnote over time on a frequency counter, we can estimate the long-term frequency stability of the lasers. Here, the standard deviation over four days is 27~kHz, and no significant drift is visible on the measurement.}
\label{fig:barbus}
\end{figure}

To amplify the lasers, we have developed custom Erbium-Doped Fiber Amplifiers (EDFA) that deliver an output of approximately 500~mW. The power efficiency of these EDFAs has been optimized to around 10 \% and their Amplified Spontaneous Emission kept to a low level, with a noise figure below 6~dB. After amplification, we use waveguide PPLN crystals to convert light from 1560 to 780~nm. For each slave laser, we obtain a power of approximately 250~mW at 780~nm, corresponding to a conversion efficiency of the order of 50\%. The two lasers are then combined in the same fiber using a polarization multiplexer and injected in a custom-made fibered AOM that sets the total output power and drives the Raman pulses. Light is finally sent in a polarization-maintaining fiber to the sensor head. At the fiber output, the power in each wavelength is approximately 150~mW and polarization extinction ratio is higher than 20~dB. We have characterized the linewidth of the laser by recording a beatnote between the gravimeter and a similar independent laser system (see Methods). We find a lorentzian linewidth of 12~kHz (Fig. \ref{fig:barbus}), which is low enough not to limit atomic cooling and detection.

\subsection*{Sensor head}

As the sensor head is intended to be installed in the field, its overall weight, compactness and ease of manipulation are essential. In this respect, we use a hollow pyramidal reflector inspired by the one described by Bodart et al.\cite{bodart_2010} to implement all the functions required by the measurement sequence with a single laser beam, leading to a significant reduction in volume, complexity and risk of misalignment (Fig. \ref{fig:visuel_AQG}).

The optical characteristics of the reflector are critical for the performance of the instrument, both in terms of polarization and wavefront aberrations\cite{trimeche_2017}. The inner faces of the reflector are coated for maximum reflection at $45^{\circ}$ and for equal phase-shift between the two orthogonal polarizations. Therefore, reflections of the single circularly-polarized beam onto the four inner faces of the pyramid create the required polarization configuration for the magneto-optical trap and molasses inside the pyramid. In addition, below the pyramid the retro-reflected laser beam can drive counter-propagating Raman transitions in a $\sigma^+ / \sigma^+$ or $\sigma^- / \sigma^-$ configuration to create the matter-wave interferometer. This polarization configuration is also suitable for fluorescence detection. The wavefront of the retroreflected beam has been characterized in detail and its quality is better than $\lambda /80$ (peak-to-valley, 780~nm) in the center region used for the atom interferometer.

The vacuum chamber is protected from external magnetic fields by two layers of mu-metal shields. A high-performance accelerometer is attached to the top of the vacuum chamber in order to implement an active compensation of vibrations and to make the instrument robust against seismic noise without the need of an isolation device\cite{merlet_2009,lautier_2014}. Two tiltmeters are also included in the sensor head to ensure the AQG is perfectly vertical, and are used to measure any long-term angular drift. Finally, a barometer is used to monitor the atmospheric pressure and correct its effect on the measurement. Because gravity gradients are of the order of 3000~nm.s$^{-2}.\mathrm{m}^{-1}$, it is important to know the effective measurement height of the instrument. On the AQG this is, on average, 55~cm above ground level. The measurement height above the leveling tripod is precisely known by design to be 48.8~cm and the height of the tripod above the ground can be measured with a precision of 1~mm during the installation of the instrument.

\subsection*{Instrument control}
Field conditions require the instrument to be stable and simple to operate. The control software of the AQG takes these constraints into account. All the operations required to start the gravimeter are automated: the software automatically locks the lasers to a predefined setpoint, turns on the EDFAs and starts the measurement sequence. Monitoring of several parameters, such as optical powers and lock stability, has been implemented and corrections are automatically applied when necessary. We have demonstrated that the laser and electronics system can run continuously for several months without breaking out of lock. At the start of the measurement sequence, the software checks additional parameters such as atom number and laser intensity, and selects the central fringe of the interferometer that gives the initial value of $g$ required to initiate the gravity measurement (see Supplementary Methods). By monitoring critical parameters over time, the software is able to detect if the instrument is unstable or requires attention.

In addition to these features, the software also calculates tilts, atmospheric pressure, vertical gravity gradients, polar motion and quartz oscillator frequency drifts. Earth tide and ocean loading corrections are also implemented, by calling a TSoft routine \cite{van_camp_2005}. Several display options are available to check the quality of the measurement in real time. Remote access to the gravimeter is possible using an internet connection, both to retrieve data and to control the instrument.

\section*{Results}
\subsection*{Scale factor verification}

The value of gravity is computed from the effective wavevector and the frequency chirp as indicated in equation (\ref{eq:g}). Both of these parameters contribute to the scale factor of the gravimeter and to the final precision of the measurement. We have measured their mean values and long-term stability at the level of a few $10^{-10}$.

The frequency chirp $\alpha$ is generated using a compact microwave synthesizer with a quartz reference oscillator. The absolute value and long-term stability of the output frequency can be calibrated using the gravimeter itself, by operating it as an atomic clock. At given time intervals, the gravity measurement is paused for approximately 30~s and the instrument switches from a Raman to a Ramsey $\pi/2 - \pi/2$ sequence where microwave pulses are driven by the microwave synthesizer. The resulting signal allows us to measure the frequency of the oscillator with an uncertainty lower than 1 part in $10^{10}$. We correct the value of $\alpha$ and $g$ accordingly, making the residual contribution lower than 1~nm.s$^{-2}$.

To ensure an absolute calibration of the laser frequencies, the system is referenced to a Rubidium optical transition at a frequency $f=c/\lambda\approx 384$~THz using saturated absorption spectroscopy. Therefore, measuring gravity with a long-term relative stability better than $10^{-9}$ requires the laser frequencies to be known with an uncertainty below 384~kHz. To characterize laser frequency stability, we use the same setup as for the linewidth measurement (see Methods). We find that the long-term frequency stability is lower than 30~kHz rms over several days (Fig. \ref{fig:barbus}), and that the effective wavector is both accurate and stable enough for the operation of the gravimeter.

\subsection*{Mitigation of external effects}

\begin{figure}[ht]
\centering
\includegraphics[width=\linewidth]{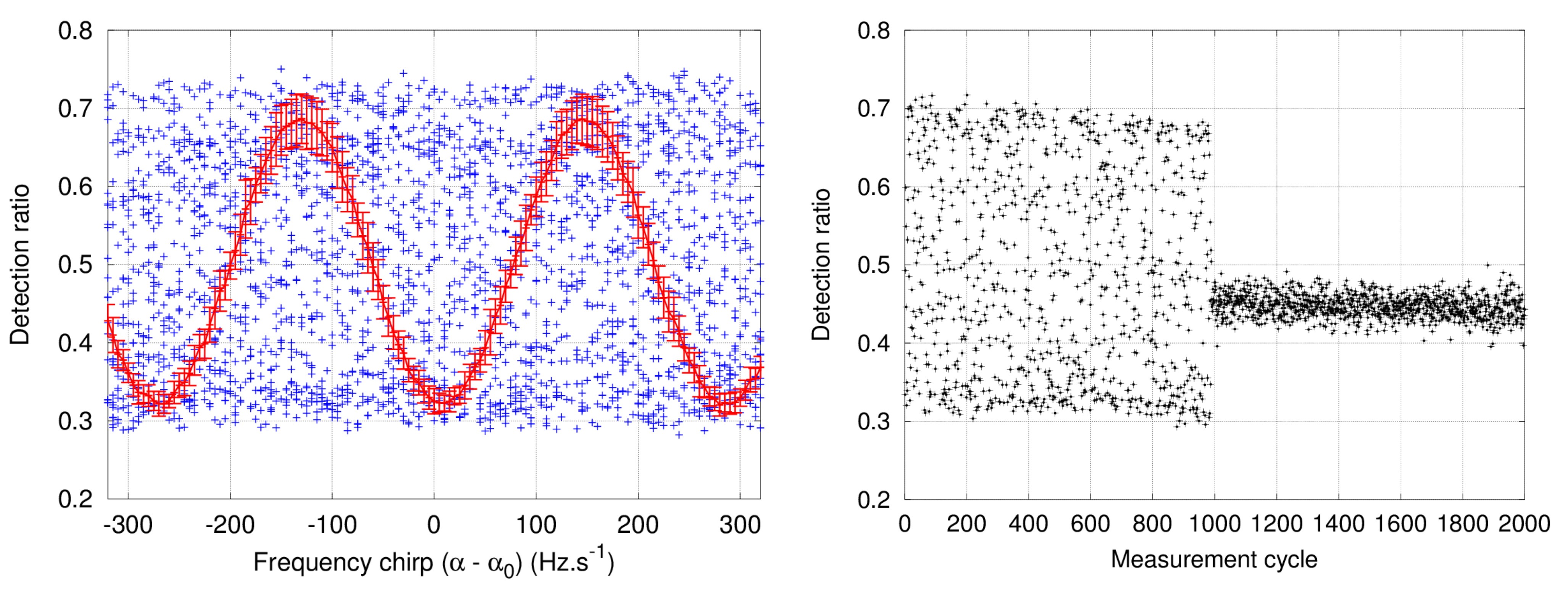}
\caption{Active compensation of vibrations. Left: atom interferometer fringes scanned by varying the Raman chirp $\alpha$. Blue crosses: without active compensation. Red line: with active compensation. Error bars correspond to the standard deviation over 20 measurements. Right: the gravimeter is operated close to mid-fringe and we measure the ratio of atoms in each internal state. During the first 1000 cycles, the compensation is turned off. Phase noise due to vibrations is 2.3~rad~rms, and the interference signal is washed out over several fringes. During the last 1000 cycles, active compensation is turned on and the vibration phase noise is greatly reduced to 36~mrad~rms, allowing the interferometer to remain close to mid-fringe.}
\label{fig:comp_vib}
\end{figure}

In order to avoid a degradation of the instrument's sensitivity by seismic noise, we have implemented an active compensation of vibrations, originally described by Merlet et al.\cite{merlet_2009} and Lautier et al. \cite{lautier_2014}. A high-performance classical accelerometer is attached to the top of the vacuum flange that supports the pyramidal reflector. This mechanical structure is very rigid by design so that the recorded signal is not distorted by resonances or deflections. The signal is filtered, digitized and weighed by the acceleration transfer function of the atom interferometer in real time (see Methods). Just before the last pulse of the interferometer a phase correction corresponding to the integrated acceleration noise is applied to the Raman lasers. The value of phase correction can be stored so that no information is lost in the process. In our laboratory in Talence (France), there is a high level of vibration noise. The integrated seismic noise typically corresponds to phase shifts of 2.3~rad rms (20~rad peak-to-peak), meaning the interference fringes are completely blurred. Using this real-time compensation, we recover the fringes and keep the residual acceleration noise to a level of approximately 36~mrad rms (250~mrad pk-pk), corresponding to a rejection factor higher than 60 (Fig. \ref{fig:comp_vib}). This techniques allows the AQG to perform sub $10^{-9}g$ measurements even in these noisy conditions.

The two high-precision tiltmeters are used to measure the tilts of the instrument in the horizontal plane and derive the vertical angle. An initial absolute calibration of the tiltmeters has been performed with the gravimeter itself (see Methods). This calibration ensures that the angle is known with an uncertainty better than 10~$\mu\mathrm{rad}$, which keeps the error on the value of $g$ below 10~nm.s$^{-2}$. The tiltmeters are first used during the installation phase to make sure the gravimeter is vertical. When the instrument is running, tilt values are continuously monitored and the corresponding correction is applied to $g$ so as to correct any long-term deviation from verticality and maintain the stability of the AQG below 10~nm.s$^{-2}$. Similarly, we use the barometer to measure atmospheric pressure variations with an accuracy better than 1~hPa. Since pressure admittance\cite{hinderer_2014} is of the order of $-3$~nm.s$^{-2}.\mathrm{hPa}^{-1}$, this is sufficient to ensure that the residual effect due to the barometer is lower than 10~nm.s$^{-2}$.

\subsection*{Sensitivity and stability measurements}

We have operated the AQG-A01 in several locations to validate its operability and transportability. In this section we discuss measurements obtained in two locations with different levels of vibration noise. The first one is the laboratory of Muquans, located in Talence (suburb of Bordeaux, France). This site features a high level of microseismic noise due to its proximity to the ocean and its location on the second floor of an inner-city building constructed on sediments. The second site is the Larzac observatory in the south of France\cite{jacob_2010,fores_2017}. This site is dedicated to hydrology studies and local gravity has been measured there by absolute and relative gravimeters on a regular basis since 2006. The Larzac observatory has a very low level of high-frequency vibration noise (i.e.~above 1~Hz) compared to Talence. A comparison of noise levels measured with the accelerometer of the AQG in both locations is shown in the Methods section. In both cases, gravity data are acquired by the AQG at a rate of approximately 2~Hz. These data are then averaged to improve statistical uncertainty, and corrected for tilt variations, drifts of the microwave oscillator, and atmospheric pressure using the admittance. Gravity data are also corrected by a synthetic tide, with parameters determined by analyzing a 9-month recording by a CG5 relative gravimeter in Talence, and a 3-year recording performed by a superconducting gravimeter in Larzac.

\begin{figure}[ht]
\centering
\includegraphics[width=\linewidth]{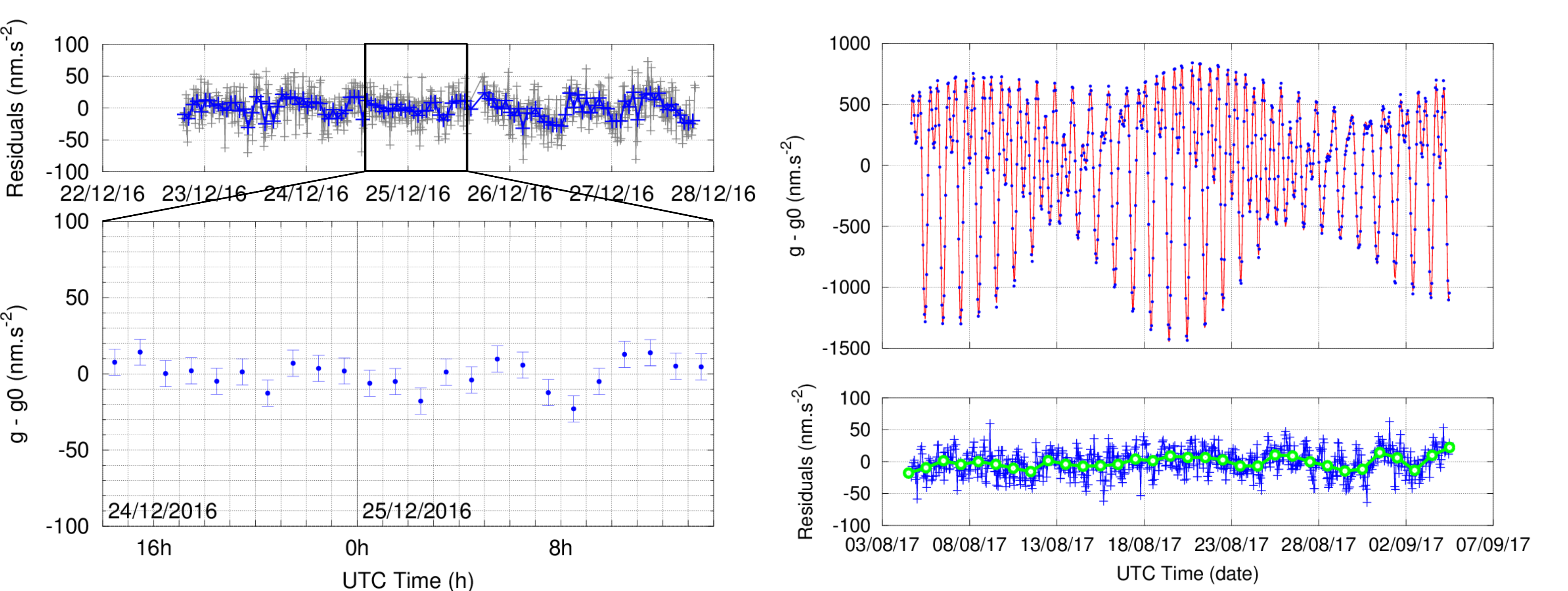}
\caption{Gravity measurements in Talence (left) and in Larzac (right). Gravity residuals are shown after correction for tides and atmospheric pressure variations. Grey: data averaged over 10~min. Blue: data averaged over 1~h. Green: data averaged over 1~day. Top left: gravity residuals in Talence. The standard deviation over the series is 25.2~nm.s$^{-2}$ (resp. 10.7~nm.s$^{-2}$) when data is averaged over 10~min (resp. 1~h). Bottom left: zoom on the best 24 hours of data. Error bars correspond to the value of the Allan deviation at 1~h of the series (8.5~nm.s$^{-2}$). Top right: tide model (red) and raw gravity in Larzac.  Bottom right: residuals. When data is averaged over 1~day, the standard deviation of the series is 9.4~nm.s$^{-2}$.}
\label{fig:residus}
\end{figure}

In Talence, we show the results of a 5~day measurement during a quiet period around Christmas 2016 (Fig. \ref{fig:residus}, left). With gravity data averaged over 1 hour, the standard deviation of these measurements is 10.7~nm.s$^{-2}$. During the best 24 hours, this deviation reduces to 8.5~nm.s$^{-2}$. The sensitivity of the instrument during the five days was approximately 600~nm.s$^{-2}.\mathrm{Hz}^{-1/2}$, mainly limited by the residual vibration noise (See the Methods section for the definition of the sensitivity). When the noise is higher, the sensitivity is typically 700~nm.s$^{-2}.\mathrm{Hz}^{-1/2}$.

During the summer of 2017 the instrument was setup in the Larzac observatory for a month-long continuous measurement. The AQG-A01 was installed on a dedicated concrete pillar directly built on the bedrock. The best sensitivity achieved by our instrument in Larzac is 500~nm.s$^{-2}.\mathrm{Hz}^{-1/2}$ and is mainly limited by imperfections in the compensation of vibrations. During the measurement campaign of 2017, the gravimeter was operated at a slightly lower sensitivity of 750~nm.s$^{-2}.\mathrm{Hz}^{-1/2}$ due to a decrease of the number of atoms loaded in the interferometer. This issue has since been resolved since, and we estimate that the instrument can now operate continuously with the nominal atom number and sensitivity for several years. We show that the instrument was able to continuously measure gravity for one month without interruption (Fig. \ref{fig:residus}, right). When data are averaged over 1 day (Fig. \ref{fig:residus}, bottom right, green circles), the standard deviation over the series reaches 9.4~nm.s$^{-2}$. As gravity was simply corrected for atmospheric effects and tides, residual fluctuations on the scale of 10~nm.s$^{-2}$ or less could come from instrumental effects, from imperfections in the correction of pressure or from hydrological and geophysical effects not predicted by the tide model. There is no measurable long-term drift in the data. This was confirmed by two measurements carried out with FG5$\#228$ absolute gravimeter on July 25$^{\rm th}$  2017 and September 4$^{\rm th}$  2017. The second FG5 value is lower by 20~nm.s$^{-2}$ (within statistical uncertainty), which shows that there has been no significant long-term gravity change over this period.

\begin{figure}[ht]
\centering
\includegraphics[width=\linewidth]{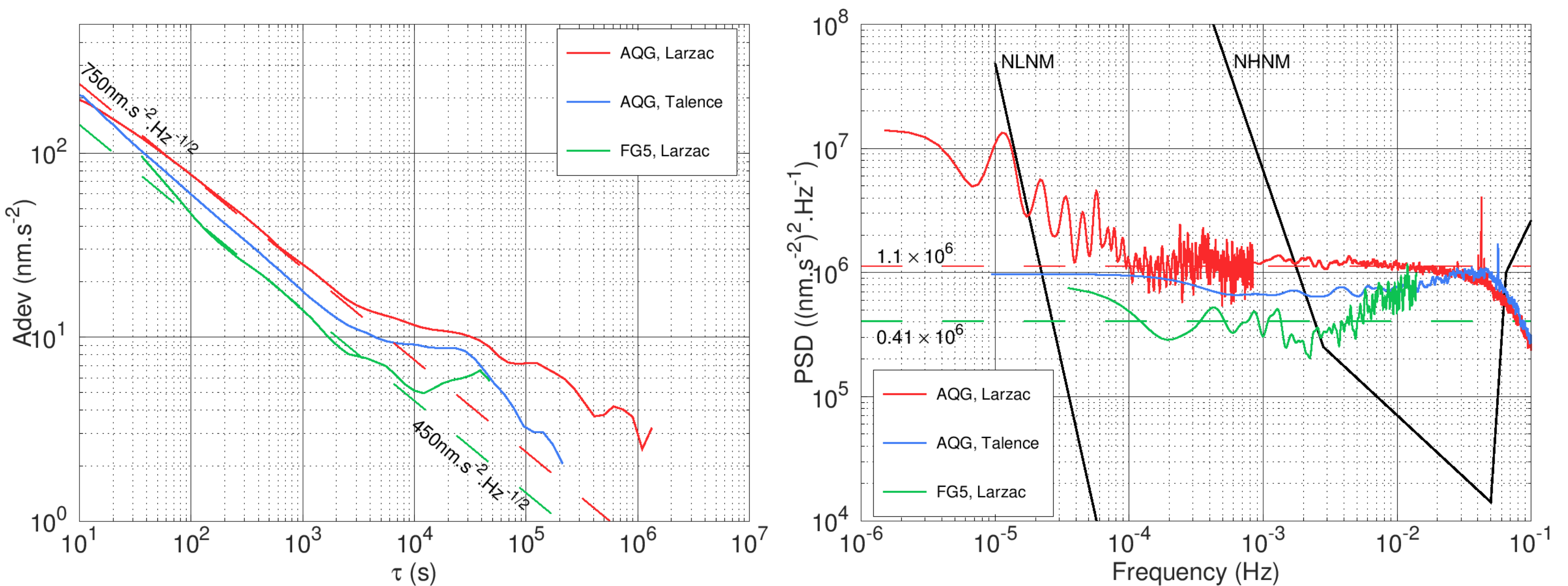}
\caption{Allan deviation (left) and power spectral density (right) of the gravity measurements with AQG-A01 in Larzac (solid red) and Talence (blue), and with FG5$\#228$ in Larzac (solid green). The effective sampling interval of the FG5 was taken as 36~s. The red (resp. green) dashed line in the Allan plot indicates a sensitivity of 750 (resp. 450)~nm.s$^{-2}.\mathrm{Hz}^{-1/2}$. This corresponds to a white noise level of 1.1 (resp. 0.41)~$\times 10^6$~$(\mathrm{nm.s}^{-2})^2.\mathrm{Hz}^{-1/2}$ in the PSD plot (see Methods). The two black lines indicate the New High and Low Noise Models from Peterson\cite{peterson_1993}. The decrease in the PSD of the AQG at frequencies higher than 0.05~Hz comes from the bandwidth of the servo-loop used to lock the frequency chirp (approximately 4~s). The sensitivity of the AQG was slightly lower in Larzac because of a decrease of the number of atoms loaded in the interferometer.}
\label{fig:Allan_PSD}
\end{figure}

The performance of the AQG during these two measurements can be characterized using the Allan deviation and the Power Spectral Density (PSD), as shown in Fig. \ref{fig:Allan_PSD}. We also show for comparison the results from the 32~hour long measurement carried out with FG5$\#228$ on September 4$^{\rm th}$ 2017. The measurement consisted of 32 sets of 100 drops. Drops were separated by 10~s and sets by 1~h, with a resulting set-scatter of 9~nm.s$^{-2}$. To plot the Allan deviation and PSD, we assumed that all the FG5 drops were continuous, with an effective measurement rate of 36~s. This reflects the performance of the instrument with the chosen settings, which include significant dead times. The results show that although the FG5 has a slightly better short-term sensitivity of approximately 450~nm.s$^{-2}.\mathrm{Hz}^{-1/2}$, the long-term stabilities of the two instruments reach similar levels, below 10~nm.s$^{-2}$. If the FG5 had measured continuously every 10~s, the sensitivity would have been improved by a factor $\sqrt{3.6} \approx 1.9$ (reaching approximately  240~nm.s$^{-2}.\mathrm{Hz}^{-1/2}$) and the PSD would have been lower by almost 6~dB. This means that despite having an intrinsically lower single-shot sensitivity, the higher repetition rate of the AQG makes averaging more efficient. At timescales longer than $10^4$~s, the data are no longer described by white noise. However, in the three datasets presented here, no significant low-frequency drift is visible. This is mainly due to the fact that absolute instruments do not experience drifts. This effect is also enhanced because the tide models we use on both sites are precise enough at these frequencies, and because there were no measurable hydrological or geophysical events during our measurements. Over the whole spectrum, the PSDs and Allan deviations show that the performances of both instruments are comparable with values typically obtained with absolute gravimeters \cite{van_camp_2017,van_camp_2005b,freier_2016}. Furthermore, despite the high level of vibrations in Talence, the AQG can be operated without significant degradation of its performance.

\subsection*{Repeatability assessment}
\begin{figure}[ht]
\centering
\includegraphics[width=\linewidth]{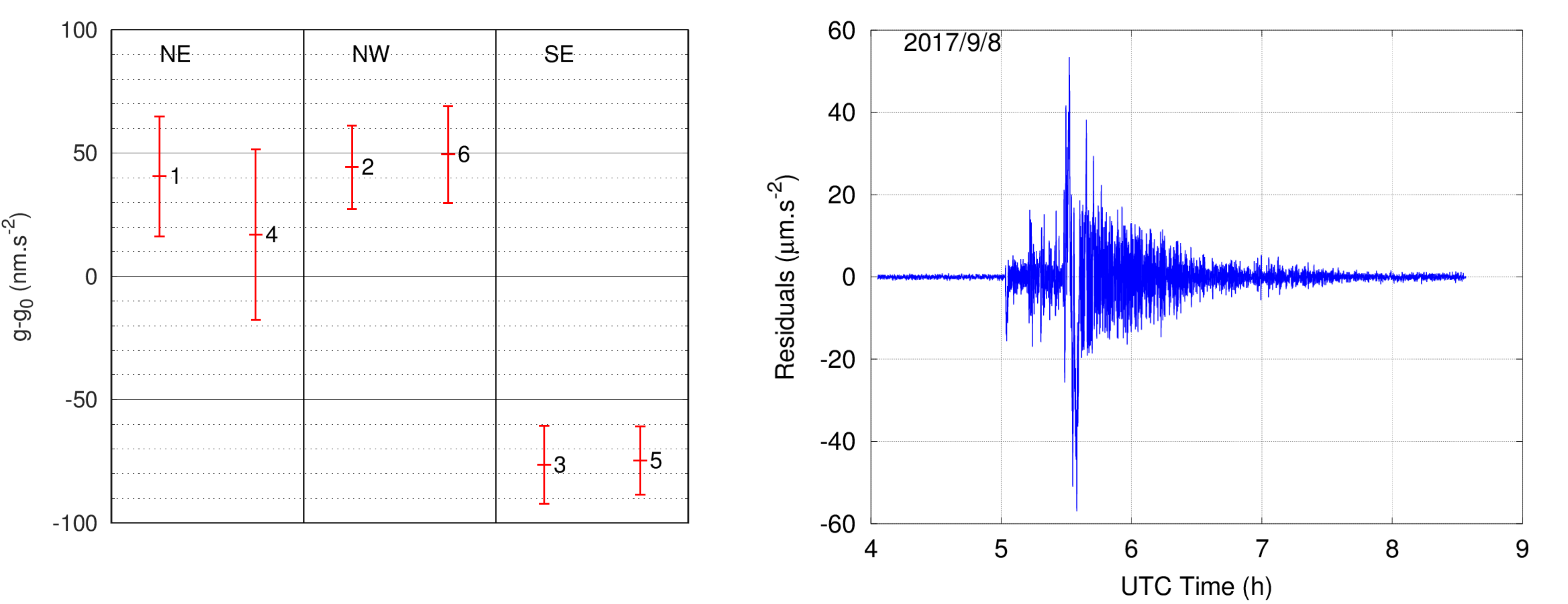}
\caption{Left: results of the repeatability measurement. The numbers next to the points indicate the order in which data were recorded (see Table \ref{tab:larzac} for details and numerical values). Right: gravity residuals measured in the Larzac observatory during the 8.1~magnitude earthquake in Mexico on September 8$^{\rm th}$, 2017. Each point corresponds to 1 measurement cycle (approx. 500~ms). Because seismic signal occurs at frequencies that are close to the high-pass cutoff of the vibration correction (see Methods), there are significant phase shifts on the acceleration signal, meaning that the resulting AQG signal is not an accurate representation of the seismic waves.}
\label{fig:Earthquake}
\end{figure}

After the one-month measurement in Larzac, we moved the sensor head of the AQG six times within the Larzac observatory to measure gravity on three pillars labeled North-East (NE), North-West (NW) and South-East (SE), to estimate the repeatability of the instrument when returning to a given location. Each pillar was measured twice, in the conditions indicated in Table \ref{tab:larzac}. On each pillar, the difference between the two measurements is within statistical uncertainty without any measurable offset (Fig. \ref{fig:Earthquake}, left).

\begin{table}[ht]
\centering
\begin{tabular}{|l|l|l|l|l|l|}
\hline
Measurement & Pillar & Date & Duration (h) & $g - g_0$ (nm.s$^{-2}$) & Statistical uncertainty (nm.s$^{-2}$)\\
\hline
\hline
1 & NE & 6 Sept 2017 & 124 & 40.5 & 24.3\\
\hline
2 & NW & 11 Sept 2017 & 20 & 44.3 & 16.9\\
\hline
3 & SE & 12 Sept 2017 & 23.5 & -76.5 & 15.8\\
\hline
4 & NE & 13 Sept 2017 & 7.5 & 16.9 & 34.5\\
\hline
5 & SE & 13 Sept 2017 & 18.5 & -74.7 & 13.9\\
\hline
6 & NW & 14 Sept 2017 & 18 & 49.4 & 19.6\\
\hline
\end{tabular}
\caption{\label{tab:larzac}Measurement conditions and results of the AQG repeatability assessment, sorted in chronological order. For convenience, measurements are expressed relative to $g_0$, which is taken arbitrarily as the average of the six measurements. The statistical uncertainty indicated in the table and used for the error bars in Fig. \ref{fig:Earthquake} is the standard deviation of the corresponding dataset, calculated after averaging the data over 1~h. The difference between the two measurements for each pillar is $16.9 - 40.5 = -23.6$~nm.s$^{-2}$ for NE, 5.1~nm.s$^{-2}$ for NW and 1.8~nm.s$^{-2}$ for SE.}
\end{table}

During this measurement campaign, the AQG also recorded several earthquakes, including the 8.1~magnitude event that occurred in Mexico on September 8$^{\rm th}$, 2017. Although the noise on gravity measurements was higher during this period, the instrument itself was not affected and did not require any specific supervision, illustrating the immunity of the AQG to external perturbations. Furthermore, the high repetition rate (2~Hz) allows for significant sampling of the accelerations caused by seismic waves, indicating that the mean gravity value is only weakly affected and that seismic signature is very clear and can be easily removed in post-processing (Fig. \ref{fig:Earthquake}, right).

\section*{Discussion}
We have presented the results obtained with an atomic quantum gravimeter (AQG-A01) based on matter-wave interferometry, enabling absolute measurements of Earth's gravity. By accounting for all operational constraints from the early stages of the design, we were able to implement several innovative developments that allowed for a mobile instrument. It is highly sensitive, stable, and compatible with a daily operation by geophysicists either for repeated short-term or continuous long-term measurements. We were able to demonstrate a high level of performance for such a compact system by achieving a continuous operation over one month with a long-term stability better than 10~nm.s$^{-2}$. A preliminary study also shows that the repeatability of the instrument is at the level of 10~nm.s$^{-2}$. As systematic effects and accuracy can only be precisely evaluated on an instrument that is stable, we are now working on their quantification in order to establish a complete accuracy budget. Our target is to characterize all the systematic effects so that the absolute value of $g$ is known with an uncertainty better than 50~nm.s$^{-2}$ and can be measured with a repeatability and stability of 10~nm.s$^{-2}$. Once a full assessment has been made, we will test it by comparing  our instrument with other absolute gravimeters.

The operation of the gravimeter relies on the use of a classical accelerometer that is utilized to filter out seismic noise. This allows the instrument to be installed directly on the ground without any vibration damping, which is a clear advantage in terms of size, weight and installation time. It also means that our AQG is robust to acceleration noise and can be used even in the event of earthquakes. Furthermore, the raw signal from the classical accelerometer, which can be recorded and stored, provides access to high-frequency accelerations (typically above 0.1~Hz), which could prove relevant in applications where these signals give complementary information to that of the gravimeter.

After years of pioneering laboratory developments and proof-of-principle experiments, the results presented here represent a significant advance in cold-atom-based technology and the use of matter-wave interferometry in geophysics. Simplified operation and continuous data acquisition capability over long periods of time open new pathways for long-term absolute gravity monitoring. The AQG can also improve and accelerate gravity surveys by removing the need for repeated measurements that are necessary to eliminate the drift of relative meters \cite{jacob_2010}.

We believe that the limits have not yet been reached with the instrument presented here \cite{farah_2014,freier_2016}. Short-term sensitivity is currently limited by imperfections in the compensation of vibrations. We are working on new low-noise electronics and real-time data processing to improve this point and to increase the SNR of the instrument. Ongoing technological developments will also reduce the overall size, weight and power consumption of the gravimeter. This includes increasing the level of integration of the electronics and simplifying the system and the connection between the various parts. Thermal management will also extend its operating temperature range from $18-30^\circ$C to $0-40^\circ$C and make it compatible with outdoor operation. To facilitate surveys, the instrument will be battery-powered, with the possibility to keep it turned on during transportation from one site to the next. With these improvements, the AQG could be used for geophysical measurements that often take place in uncontrolled environments.

\section*{Methods}

\subsection*{Atomic temperature measurement}
To estimate the temperature of the atomic cloud, we use Raman spectroscopy. After cooling and preparation, the atoms are dropped. We drive a Raman transition with a 80~$\mu \mathrm{s}$ pulse and detect fluorescence at the bottom of the chamber. The resulting profile is the convolution of the atomic velocity distribution by the Raman pulse width. Assuming the latter to be negligible we measure the full width at half maximum (FWHM) of the distribution and derive the corresponding temperature, below 2~$\mu$K.

\subsection*{Vibration acquisition and compensation}
The signal from the classical accelerometer (Nanometrics Titan) is filtered by a bandpass filter with cut-off frequencies $f_\mathrm{hp} = 0.05$~Hz and $f_\mathrm{lp} = 1$~kHz. The high-pass filter removes any long-term drift of the accelerometer and makes sure that the mean gravity value only comes from the atomic measurement. $f_\mathrm{hp}$ has been chosen much lower than the cycle frequency of the AQG (2~Hz), so that there is no significant phase shift of the signal at 2~Hz. The low-pass filter essentially eliminates high-frequency electronic noise before the signal is digitized. Since the atom interferometer behaves as a second-order filter with a low-pass cut-off frequency of $1/2T = 8.3$~Hz, $f_\mathrm{lp}$ is chosen significantly higher. This way, we make sure the electronic low-pass filter does not impact the measurement at frequencies where the atom interferometer has a residual sensitivity.

The filtered signal is digitized and weighted in real-time by the transfer function of the atom interferometer\cite{cheinet_2008,geiger_2011}. The acceleration response function has a simple triangle shape so the real-time calculation is straightforward. It has been demonstrated that a convenient and yet robust way of taking into account the response function of the system is to introduce a delay on the application of the sensitivity function\cite{lautier_2014}. This delay is simply calculated by optimizing the signal-to-noise ratio of the instrument. Less than 1~ms before the last Raman pulse, the calculated correction is applied to the phase lock loop of the Raman lasers to compensate the vibration phase shift.

\begin{figure}[ht]
\centering
\includegraphics[width=0.5\linewidth]{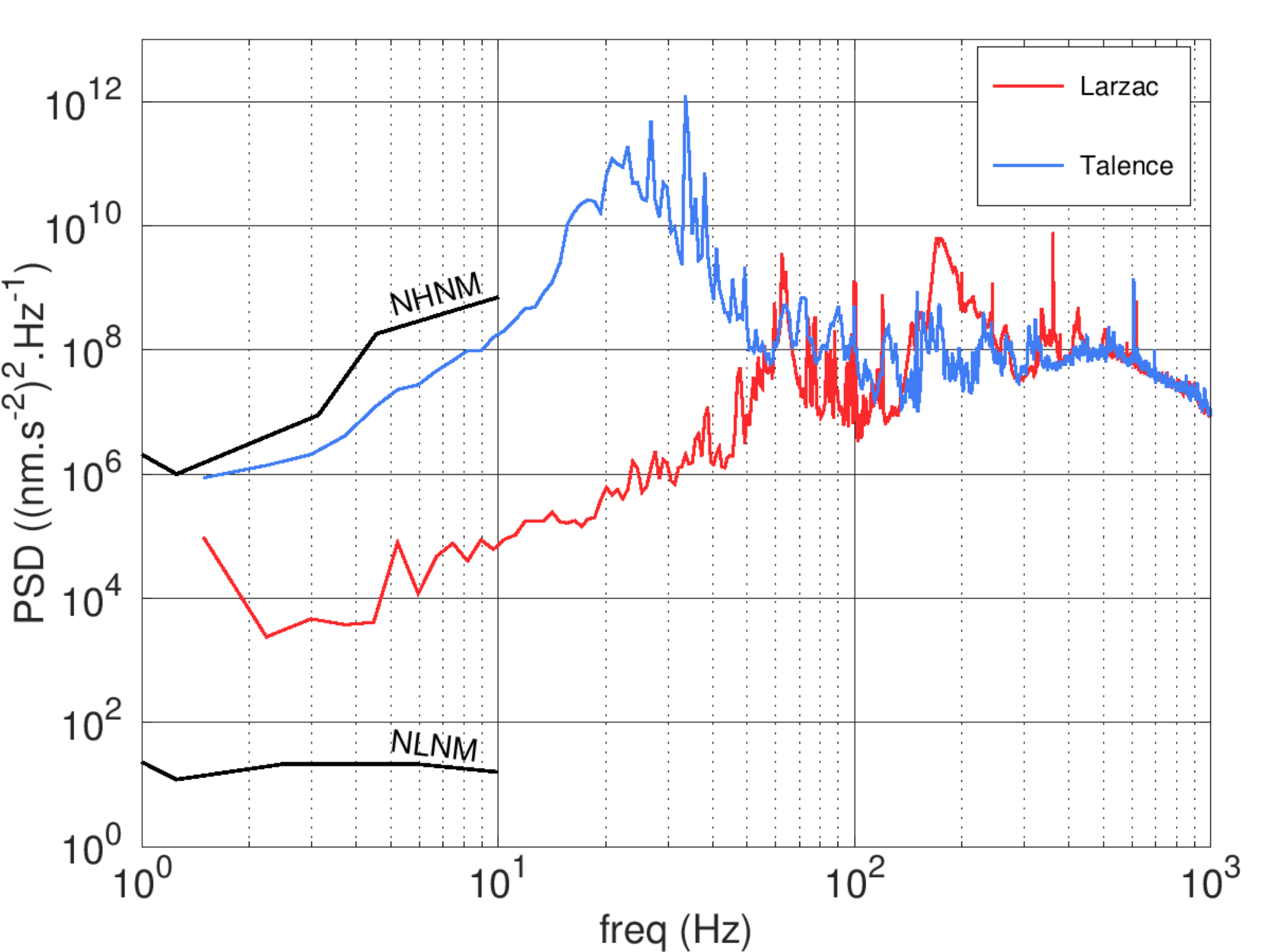}
\caption{Vibration spectra measured with the classical accelerometer attached to the AQG in Talence and Larzac, France.}
\label{fig:vibrations}
\end{figure}

Acceleration data can also be used to perform an analysis of the acceleration PSD, which can be useful to estimate the quality of a measurement location in terms of high-frequency vibrations. For example, spectra recorded in the laboratory of Muquans in Talence and at the Larzac observatory show significant differences, especially at frequencies between 2 and 60~Hz (Fig. \ref{fig:vibrations}). This can be explained by several effects such as the proximity to the Atlantic ocean (50~km from Talence) which produces microseismic noise, oscillations of the building and the fact that it is built on sediments rather than solid bedrock. The high level of human activity in Talence (second-floor of a city building with a nearby tramway line) could also be a contributor. Note that effects below $f_\mathrm{hp}$ are not measured in the data of Fig. \ref{fig:vibrations} because they are rejected by the filter. Some of them have a direct impact on the value of $g$, while others are measured and corrected by the software (e.g.~tilts and atmospheric pressure variations).

\subsection*{Linewidth and long-term frequency stability measurement}
We record the beatnote between the laser of the AQG and a similar independent laser to estimate the linewidth and frequency stability of the gravimeter laser. The two lasers are in a master / slave configurations, so they have adjustable setpoints. We typically use a frequency difference on the order of 80~MHz.

The beatnote is recorded on a fast photodiode and sent to an RF spectrum analyzer. The center of the spectrum can be fitted by a Gaussian function accounting for technical noise. The tails (more than $5 \sigma$ away from the center) are fitted by a Lorentzian function. Assuming the two lasers to be independent, the width of the fitted function is the sum of the two individual widths. Also assuming the two lasers to be identical, we estimate their linewidth as half of this. In the case of Fig. \ref{fig:barbus}, the fitted Lorentzian width is 23.2~kHz, meaning that both lasers have a linewidth of less than 12~kHz.

To measure the long-term frequency stability, the beatnote signal is sent to a frequency counter and compared to a stable RF reference. We measure the standard deviation of the beatnote frequency and expect that each of the two lasers has a stability lower than this value (27~kHz in the case of Fig. \ref{fig:barbus}). Assuming similarity and independence of the two lasers as above, this corresponds to a worst-case estimation.

\subsection*{Tiltmeter offset correction}
We measure the offset of the tiltmeters by performing an in-situ calibration that also takes into account any additional angle due to imperfect alignments between the tiltmeters and the pyramid reflector. This calibration only has to be performed once, after assembling the sensor head.

We apply known tilts in both directions to the instrument and measure the resulting value of gravity. Applied tilts typically range from 0 to 1.5~mrad, corresponding to gravity variations of approximately 10~$\mu$m.s$^{-2}$. We fit the data to find the position of the maximum, that corresponds to verticality.

The measurement sensitivity is limited by statistical uncertainties on the values of $g$, giving an estimation of the tiltmeter offsets with a precision lower than 10~$\mu\mathrm{rad}$. This is sufficient to ensure that tilts can be corrected from the final gravity measurements with a precision better than 10~nm.s$^{-2}$.

\subsection*{Sensitivity estimation}
We use the Allan deviation and PSD of the data to estimate the sensitivity and stability of the instrument \cite{riley_2008}.

The data we obtain from both the AQG and the FG5 display a classical white noise behaviour over a large frequency range. In this regime, the Allan deviation decreases in proportion with the square root of the averaging time $\tau$. In the log-log plot of Fig. \ref{fig:Allan_PSD} this behaviour is characterized by a linear decrease with a slope of $-1/2$. The sensitivity $\mathcal{S}$ of the instrument, expressed in nm.s$^{-2}.\mathrm{Hz}^{-1/2}$, is the extrapolation of the white noise behaviour to $\tau = 1$~s. This can be conveniently interpreted as the statistical uncertainty obtained after averaging data over 1~s.

In the Allan deviation plot, all three data sets exhibit a clear white-noise signature between 100 and 2000~s. The corresponding sensitivities are indicated by the dashed lines: approximately 450~nm.s$^{-2}.\mathrm{Hz}^{-1/2}$ for the FG5 and 750~nm.s$^{-2}.\mathrm{Hz}^{-1/2}$ for the AQG measurement in Larzac. In the PSD plot, this corresponds to constant levels between $5 \times 10^{-4}$ and $1 \times 10^{-2}$~Hz. The value $S$ of the PSD (expressed in $(\mathrm{nm.s}^{-2})^2.\mathrm{Hz}^{-1}$) in this white noise region is related to the sensitivity by \cite{riley_2008}
\begin{equation}
S = 2 \times \mathcal{S}^2.
\end{equation}

At longer timescales, both the PSD and Allan deviation show that the data can no longer be described as white noise, indicating instrumental drifts or long-term geophysical effects\cite{van_camp_2017}.

\subsection*{Data availability}
The datasets generated during and/or analysed during the current study are available from the corresponding author on reasonable request.

\section*{Acknowledgements}
Development of the AQG was funded through the EquipEx RESIF-CORE which is supported by a public grant overseen by the French national research agency (ANR) as part of the Investissements d’Avenir program (ANR-11-EQPX-0040). We also acknowledge support from Direction G\'en\'erale de l'Armement (GRAVITER project), R\'egion Aquitaine and Idex Université de Toulouse, IRD and CNRS (GEOMIP project). The authors acknowledge a major contribution from the Muquans team in the design, integration, characterization and optimization of the Absolute Quantum Gravimeter. They thank Franck Pereira Dos Santos and Sébastien Merlet for helpful discussions, and Baptiste Battelier for his support in the early-stage design of the instrument.

\end{document}